\documentclass[11pt]{article}
\usepackage{graphicx}
\usepackage{subfigure}

\newcommand{\BABARPubYear}    {06}
\newcommand{\BABARConfNumber} {01}
\newcommand{\SLACPubNumber} {11980}

\input babarsym

\newcommand{\kevcc}{\ensuremath{{\mathrm{\,ke\kern -0.1em V\!/}c^2}}\xspace}

\def\LcD{\ensuremath{\Lambda_{c}^{+}\rightarrow pK^-\pi^+}}

\def\babar{\mbox{\slshape B\kern-0.1em{\smaller A}\kern-0.1em B\kern-0.1em{\smaller A\kern-0.2em R}}\xspace}

\def\A31{\mbox{\tt Analysis31}\xspace}

\long\def\inst#1{\par\nobreak\kern 4pt\nobreak
    {\it #1}\par\vskip 10pt plus 3pt minus 3pt}

\begin{document}
{\pagestyle{empty}

\begin{flushright}
\babar-CONF-\BABARPubYear/\BABARConfNumber \\
SLAC-PUB-\SLACPubNumber \\
July 2006 \\
\end{flushright}

\par\vskip 2cm

\begin{center}
\Large \bf A Study of $\Xi_c(2980)^+$ and $\Xi_c(3077)^+$
\end{center}
\bigskip

\begin{center}
\large The \babar\ Collaboration\\
\mbox{ }\\
\today
\end{center}
\bigskip \bigskip

\begin{center}
\large \bf Abstract
\end{center}
We present a study of two states decaying to $\Lambda_c^+K^-\pi^+$ using the \babar\ detector at the SLAC PEP-II asymmetric-energy \epem storage rings.
We use an integrated luminosity of $288.5\invfb$ collected at the center-of-mass energy $\sqrt{s} = 10.58$~\gev, near the peak of the $\Upsilon(4S)$ resonance, plus $27.2\invfb$ collected approximately $40\mev$ below this energy.
We search for the particles $\Xi_{c}(2980)^+$ and $\Xi_{c}(3077)^+$, recently discovered by the Belle Collaboration, in their decays to $\Lambda_c^+K^-\pi^+$, where \LcD.
We find a signal with $7.0\,\sigma$ significance for the $\Xi_c(2980)^+$ state with a mass difference with respect to the $\Lambda_c^+$ of $(680.6\pm1.9\pm1.0)\mevcc$ (first error is statistical and second error is systematic).
The measured width for this state is $(23.6\pm2.8\pm1.3)\mev$, and the yield is $284\pm45\pm46$ events.
We find a signal with $8.6\,\sigma$ significance for the $\Xi_c(3077)^+$ state with a mass difference with respect to the $\Lambda_c^+$ of $(790.0\pm0.7\pm0.2)\mevcc$, a width of $(6.2\pm1.6\pm0.5)\mev$, and a yield of $204\pm35\pm12$ events.
The $\Xi_{c}(2980)^+$ is found to decay resonantly through the intermediate state $\Sigma_c(2455)^{++}K^-$ with $4.9\,\sigma$ significance and non-resonantly to $\Lambda_c^+K^-\pi^+$ with $4.1\,\sigma$ significance. 
With $5.8\,\sigma$ significance, the $\Xi_c(3077)^+$ is found to decay resonantly through $\Sigma_c(2455)^{++}K^-$, and with $4.6\,\sigma$ significance, it is found to decay through $\Sigma_c(2520)^{++}K^-$.
The significance of the signal for the non-resonant decay $\Xi_c(3077)^+\rightarrow\Lambda_c^+K^-\pi^+$ is $1.4\,\sigma$.
These results are preliminary.
\vfill
\begin{center}

Submitted to the 33$^{\rm nd}$ International Conference on High-Energy Physics, ICHEP 06,\\
26 July---2 August 2006, Moscow, Russia.

\end{center}

\vspace{1.0cm}
\begin{center}
{\em Stanford Linear Accelerator Center, Stanford University, 
Stanford, CA 94309} \\ \vspace{0.1cm}\hrule\vspace{0.1cm}
Work supported in part by Department of Energy contract DE-AC03-76SF00515.
\end{center}

\newpage
}

\begin{center}
\small

The \babar\ Collaboration,
\bigskip

%
{B.~Aubert,}
{R.~Barate,}
{M.~Bona,}
{D.~Boutigny,}
{F.~Couderc,}
{Y.~Karyotakis,}
{J.~P.~Lees,}
{V.~Poireau,}
{V.~Tisserand,}
{A.~Zghiche}
\inst{Laboratoire de Physique des Particules, IN2P3/CNRS et Universit\'e de Savoie,
 F-74941 Annecy-Le-Vieux, France }
% ...

\end{center}\newpage

\section{INTRODUCTION}
\label{sec:Introduction}

The Belle Collaboration has recently reported evidence for two new charm baryon states~\cite{Belle1,Belle2}.
These two new states have been called the $\Xi_c(2980)^+$ and the $\Xi_c(3077)^+$.
Belle finds evidence for these states at $(2978.5\pm2.1\pm2.0)\mevcc$ and $(3076.7\pm0.9\pm0.5)\mevcc$ in the $\Lambda_c^+K^-\pi^+$ invariant mass spectrum
and quotes statistical significances greater than $6\,\sigma$ for both states.
They also find significant signal for the isospin partner $\Xi_c(3077)^0$ in the $\Lambda_c^+K_S^0\pi^-$ invariant mass spectrum.

Previously known excited $\Xi_c$ baryons have only been observed in decays to lower-mass $\Xi_c$ baryons plus a pion or gamma.
These two new states decay such that the charm and strange quarks are contained in separate hadrons.
This type of decay may have implications for the internal quark dynamics of these two new states.

In the analysis described here, we find yields for $\Xi_c(2980)^+$ and $\Xi_c(3077)^+$ in their decays to $\Lambda_c^+K^-\pi^+$, where $\Lambda_c^+\rightarrow pK^-\pi^+$.
The mass differences with respect to the $\Lambda_c^+$ baryon and the widths of $\Xi_c(2980)^+$ and $\Xi_c(3077)^+$ are also measured.
The Dalitz-plot structure of the $\Xi_c(2980)^+$ and the $\Xi_c(3077)^+$ three-body decays are also studied.
The statistical significance of resonant decays through $\Sigma_c(2455)^{++}K^-$ and $\Sigma_c(2520)^{++}K^-$ are calculated,
and the signal yields for both resonant and non-resonant decays are measured.
 
\section{THE \babar\ DETECTOR AND DATASET}
\label{sec:babar}

We use $288.5\invfb$ of data collected at $\sqrt{s} = 10.58$~\gev plus $27.2\invfb$ of data collected approximately $40\mev$ below this energy. The \babar\ detector, located at the SLAC PEP-II asymmetric-energy \epem storage rings, was used to collect this data. The \babar\ detector is described in detail elsewhere~\cite{ref:babar}. The tracking of charged particles is provided by a five-layer double-sided silicon vertex tracker (SVT) and a 40-layer drift chamber (DCH). Discrimination among charged pions, kaons, and protons relies on ionization energy loss (\dedx) in the DCH and SVT, and on Cherenkov photons detected in a ring-imaging detector (DIRC). A CsI(Tl) crystal calorimeter is used to identify electrons and photons. These four detector subsystems are mounted inside a 1.5-T solenoidal superconducting magnet. The instrumented flux return for the solenoidal magnet provides muon identification.
 
For signal event simulations, we use the Monte Carlo (MC) generators JETSET74 \cite{jetset} and EVTGEN~\cite{EVTGEN} with a full detector simulation based on GEANT4~\cite{MC}.
These simulations are used to estimate the reconstruction efficiencies and detector resolutions.
The efficiencies in our analysis for finding simulated $\Xi_c(2980)^+$ and $\Xi_c(3077)^+$ decays to $\Lambda_c^+K^-\pi^+$, where $\Lambda_c^+\rightarrow pK^-\pi^+$, are roughly $10\%$.

\section{ANALYSIS METHOD}
\label{sec:Analysis}

\begin{figure}
\begin{minipage}[t]{0.48\textwidth}
\includegraphics[width=.98\textwidth]{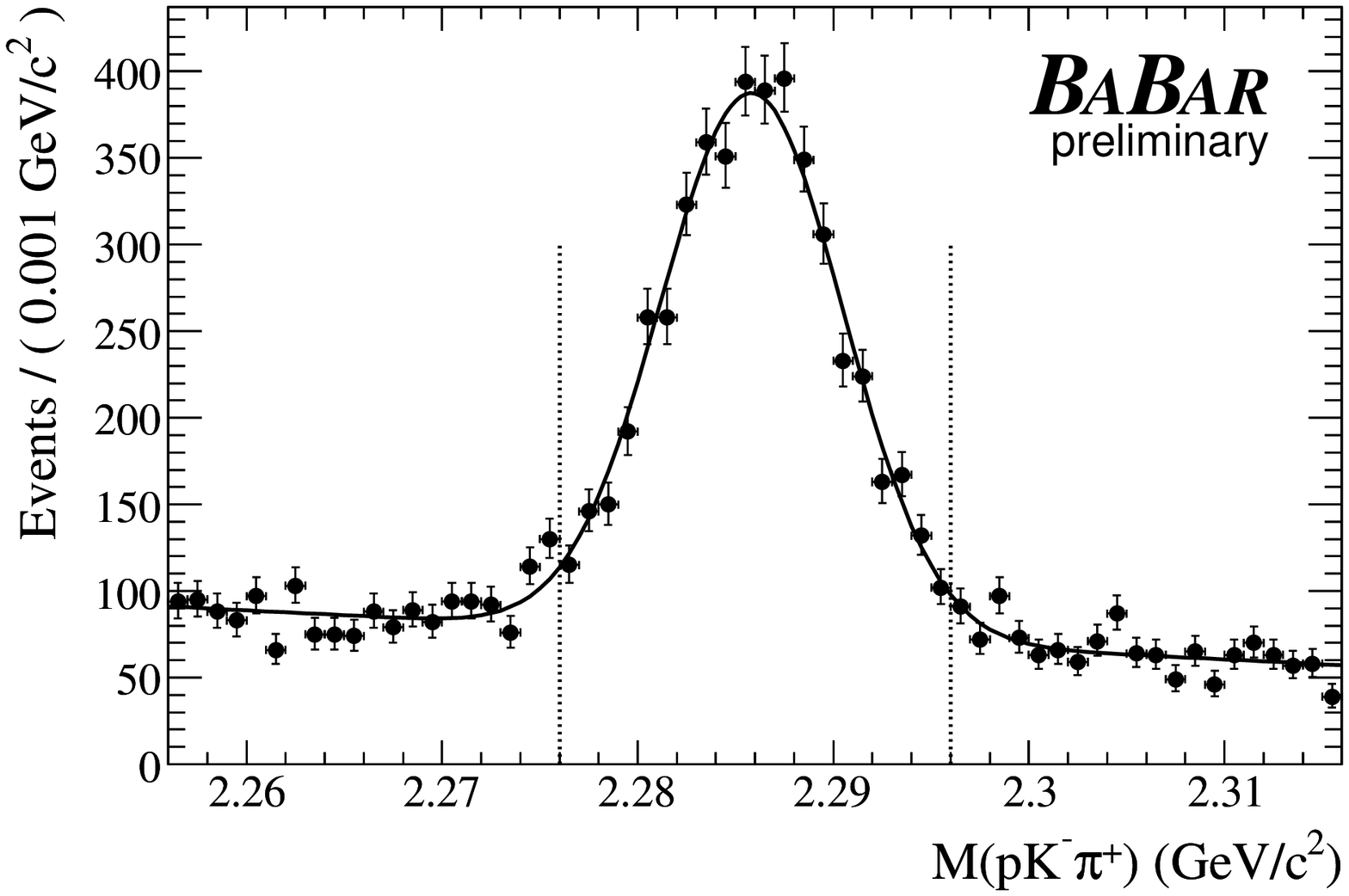}
\caption{The $pK^-\pi^+$ invariant mass distribution for signal candidates in data. The curve depicts an unbinned likelihood fit to a Gaussian plus a line. The dotted vertical lines are $\pm 10\mevcc$ ($\pm2.1~\sigma$) from the mean of the fitted Gaussian (2286\mevcc).}
\label{fg:lcrange}
\end{minipage}
\hfill
\begin{minipage}[t]{0.48\textwidth}
\includegraphics[width=.98\textwidth]{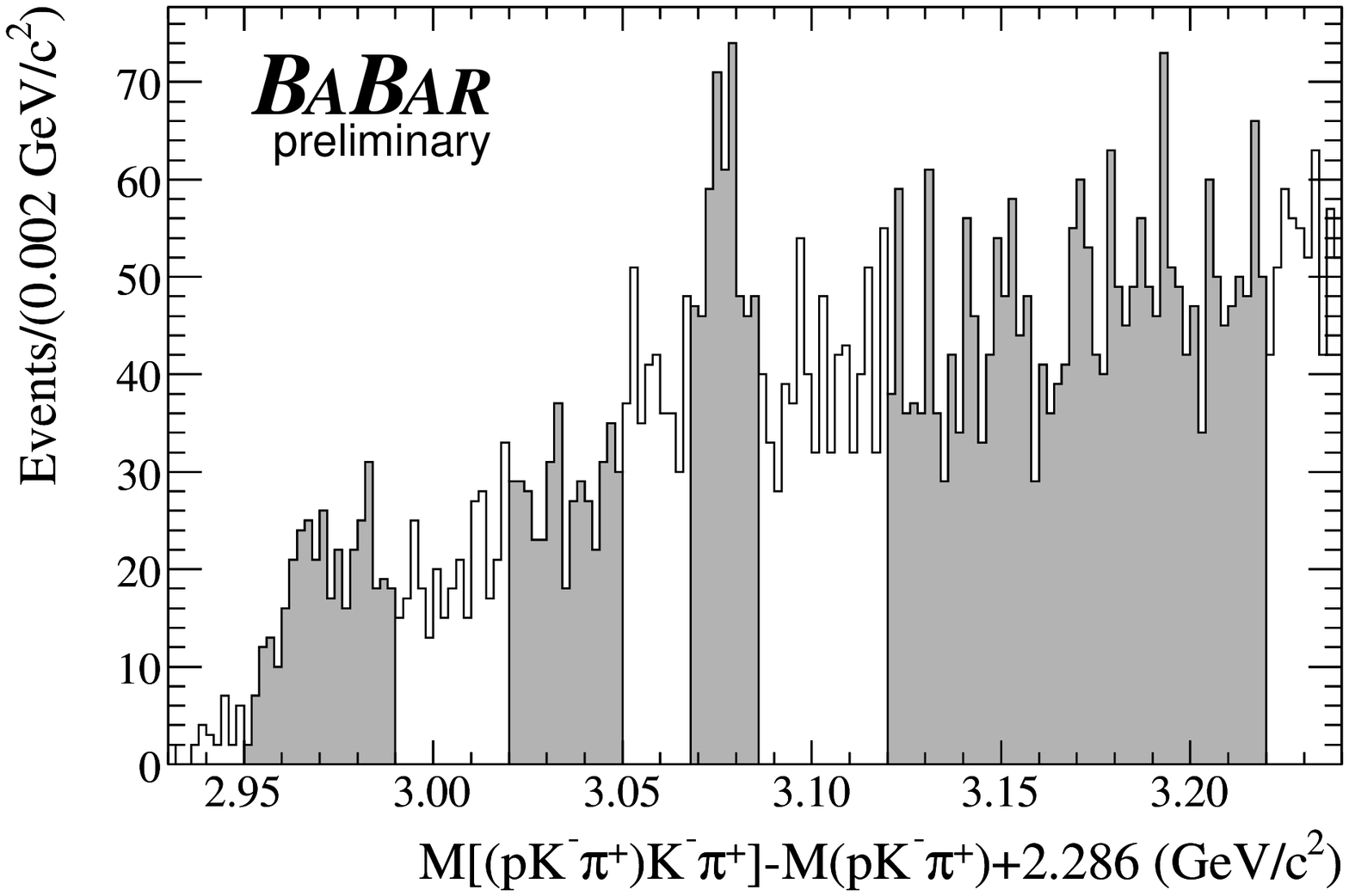}
\caption{The distribution of the invariant mass difference $M_{\Xi_c}$ for signal candidates in data. The shaded regions are used for the Dalitz plots illustrated in Figure~\ref{fig:dall}.}
\label{fg:dranges}
\end{minipage}
\end{figure}

$\Lambda_c^+$ candidates are formed from the geometrical combination of $p$, $K^-$, and $\pi^+$ tracks.
$\Xi_c(2980)^+$ and $\Xi_c(3077)^+$ candidates are formed from combining the $\Lambda_c^+$ candidates with additional $K^-$ and $\pi^+$ tracks.  
The $pK^-\pi^+$ and $\Lambda_c^+K^-\pi^+$ vertices are fit simultaneously.
The selection criteria, based on particle identification likelihood ratios, track qualities (tracks within $1.5\cm$ of the beam spot and with DCH information), and the $\chi^2$ probability for fitted vertices ($P(\chi^2)>1\%$), are designed to maximize $\epsilon/\sqrt{B}$.
The value of $\epsilon$ is the simulated signal reconstruction efficiency and $B$ is the number of background candidates in data.
The $\Lambda_c^+$ candidate mass is required to be within 10\mevcc ($2.1\sigma$) of 2286\mevcc (the fitted mean of the $\Lambda_c^+$ signal in the data) as illustrated in Figure~\ref{fg:lcrange}.
Charm hadrons carry a significant fraction of the initial energy of the charm quark, whereas random combinations of charged particles in an event form lower-energy candidates.
To take advantage of this difference, we select signal candidates that have momentum in the \epem center-of-mass frame greater than 3.0\gevc.

Any resonant substructure in a three-body decay mode will alter the signal line shape from that of a decay mode with a uniform phase-space substructure.
This is particularly important when the decay in question is near its kinematic threshold.
The resonant substructure of the $\Lambda_c^+K^-\pi^+$ combination is studied with Dalitz plots for four ranges of the invariant mass difference
\begin{equation}
M_{\Xi_c}=M[(pK^-\pi^+)K^-\pi^+]-M(pK^-\pi^+)+2.286\gevcc\,.
\end{equation}
These ranges are illustrated in Figure~\ref{fg:dranges}.
Candidates around the $\Xi_c(2980)^+$ and $\Xi_c(3077)^+$ signals are selected with the ranges $2.95\gevcc<M_{\Xi_c}<2.99\gevcc$ and $3.07\gevcc<M_{\Xi_c}<3.09\gevcc$, respectively.
The ranges $3.02\gevcc<M_{\Xi_c}<3.05\gevcc$ and $3.12\gevcc<M_{\Xi_c}<3.22\gevcc$ are used to select background candidates.

\begin{figure}\begin{center}
\subfigure{\label{aaa}\includegraphics[width=0.46\textwidth]{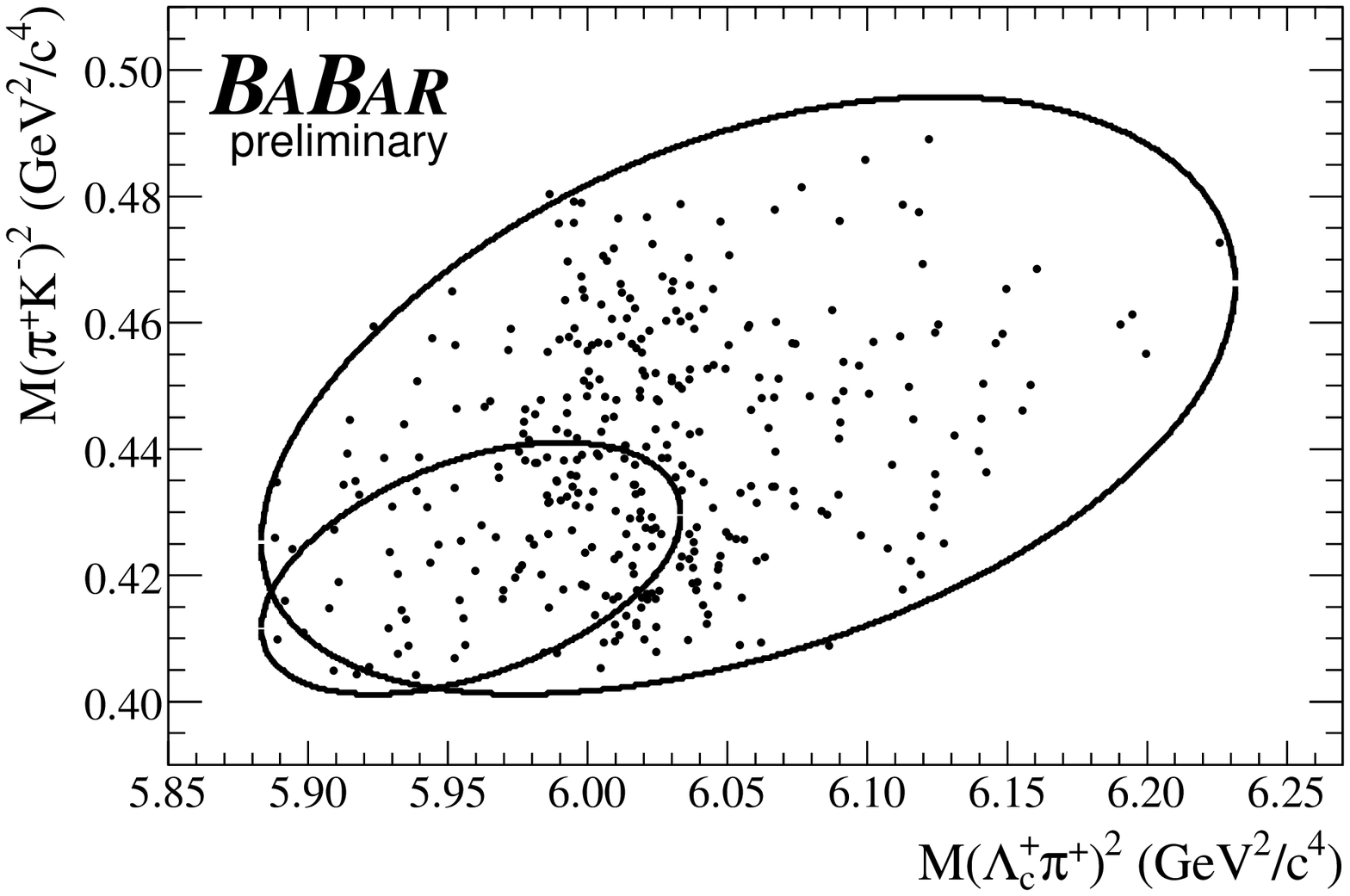}}
\hspace{0.02\textwidth}
\subfigure{\label{bbb}\includegraphics[width=0.46\textwidth]{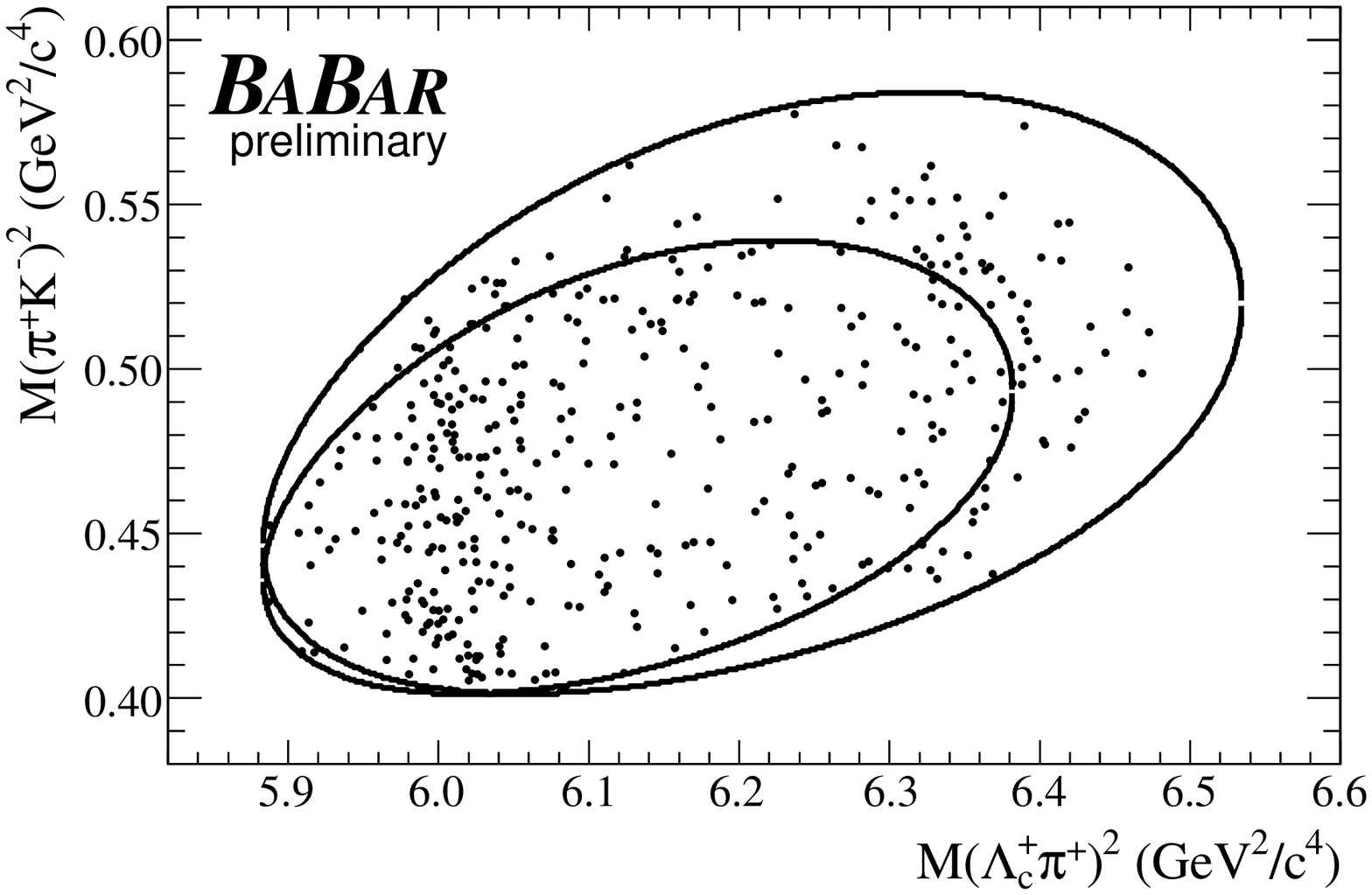}}
\hspace{0.5in}
\subfigure{\label{ccc}\includegraphics[width=0.46\textwidth]{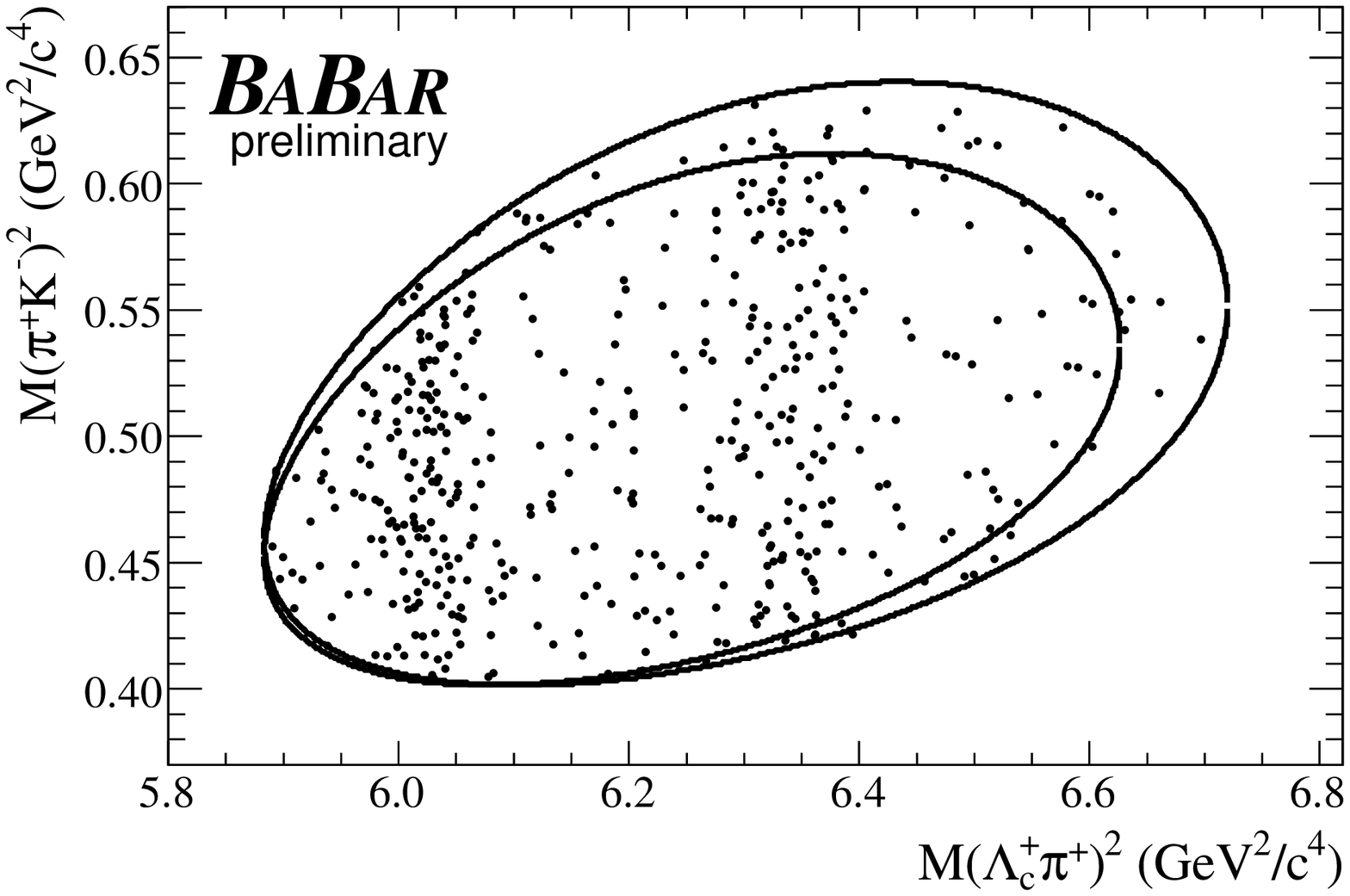}}
\hspace{0.02\textwidth}
\subfigure{\label{ddd}\includegraphics[width=0.46\textwidth]{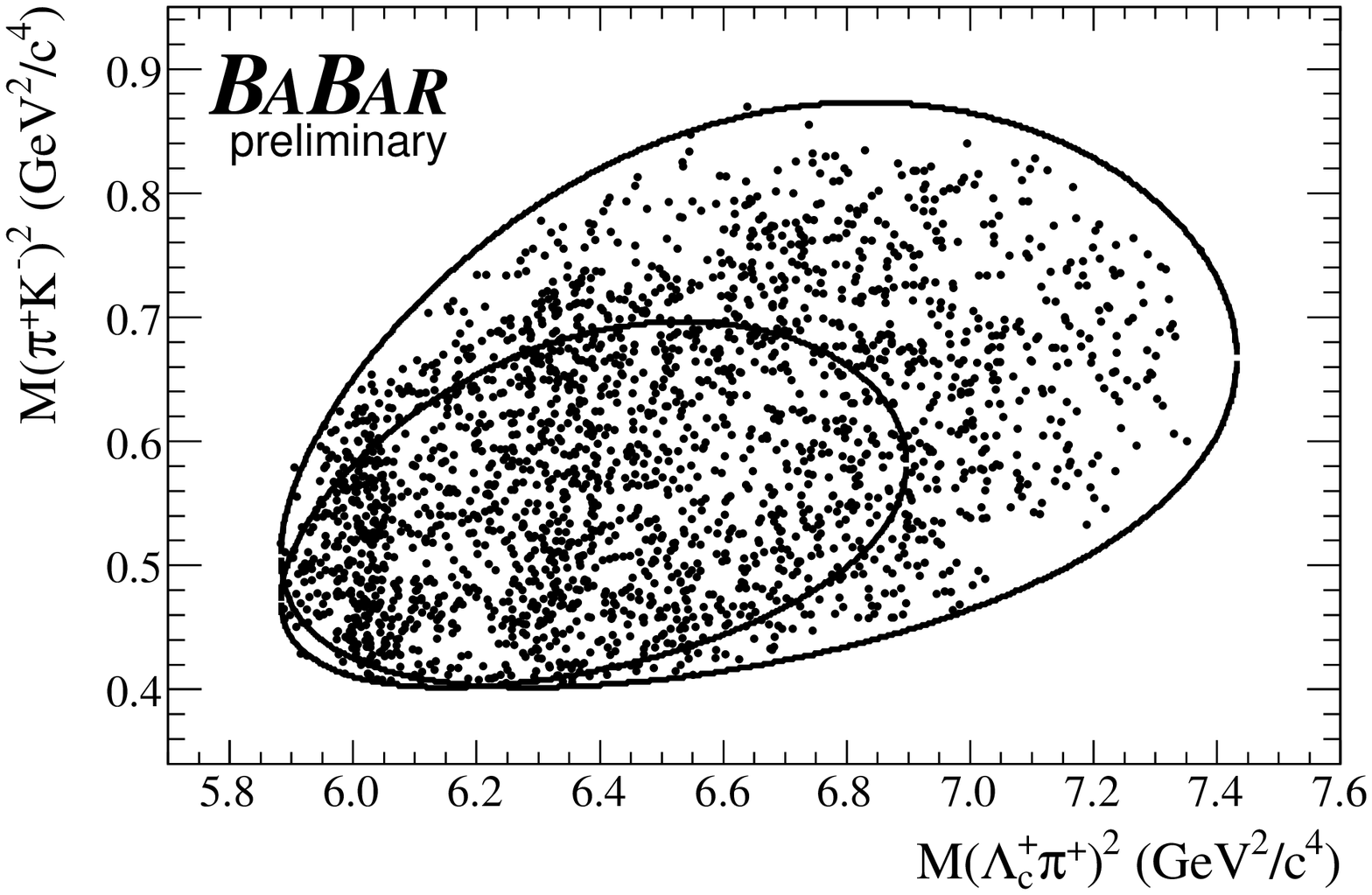}}
\caption{Dalitz plots of $M(\pi^+K^-)^2$ vs.~$M(\Lambda_c^+\pi^+)^2$. Data from the four shaded $M_{\Xi_c}$ ranges in Figure~\ref{fg:dranges} are shown in order of increasing mass from left to right, top to bottom. The two curves in each plot represent the kinematic boundaries of the Dalitz plot for $M_{\Xi_c}$ mass values at the lower and upper edges of the $M_{\Xi_c}$ range.}
\label{fig:dall}
\end{center}\end{figure}

Dalitz plots of $M(\pi^+K^-)^2$ versus $M(\Lambda_c^+\pi^+)^2$ are shown in Figure~\ref{fig:dall}
for each of the $M_{\Xi_c}$ ranges shown in Figure~\ref{fg:dranges}. In each of the four sub-figures, $\Lambda_c^+\pi^+$ resonances are visible as vertical bands at $M(\Lambda_c^+\pi^+)^{2}\sim6.02{\ensuremath{\rm \,Ge\kern -0.08em V^2\!/c^4}}$
and/or $M(\Lambda_c^+\pi^+)^{2}\sim6.35{\ensuremath{\rm \,Ge\kern -0.08em V^2\!/c^4}}$, corresponding to the $\Sigma_c(2455)^{++}$ and $\Sigma_c(2520)^{++}$, respectively.
No other significant structures are observed in the Dalitz plots.
The presence of the $\Sigma_c(2455)^{++}$ and/or $\Sigma_c(2520)^{++}$ resonances in all four $M_{\Xi_c}$ regions motivates a two-dimensional fit that can account for the effects of these intermediate resonances.

\begin{figure}\begin{center}
\begin{minipage}[t]{0.98\textwidth}
\begin{center}
\includegraphics[width=0.75\textwidth]{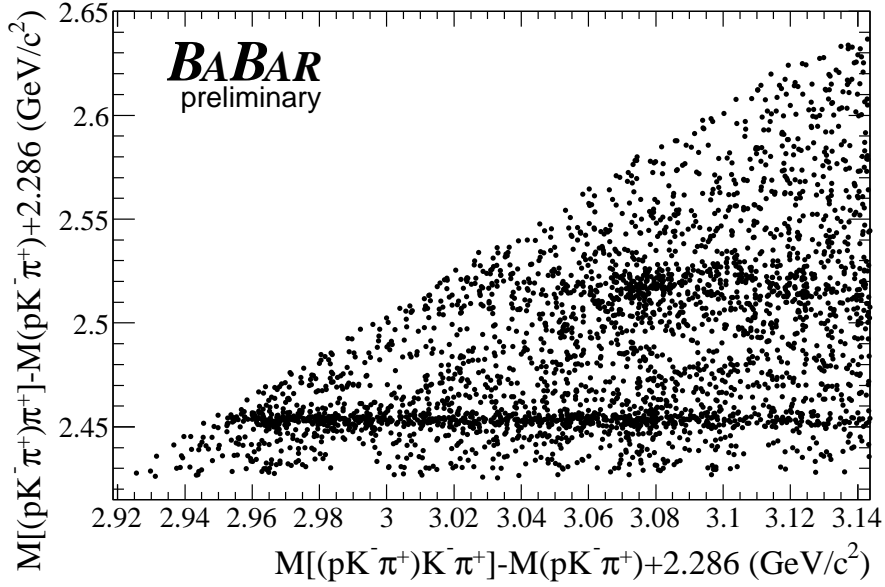}
\caption{A two-dimensional scatter plot of the $M_{\Sigma_c}$ vs.~$M_{\Xi_c}$ fit range for $pK^-\pi^+K^-\pi^+$ candidates in data. The upper and lower horizontal bands are from the $\Sigma_c(2520)^{++}$ and $\Sigma_c(2455)$ resonances, respectively.}
\label{fg:2D}
\end{center}
\end{minipage}\end{center}
\end{figure}

The data is fit in two dimensions of invariant mass difference: $M_{\Xi_c}$, as defined by Equation~1, and 
\begin{equation}
M_{\Sigma_c}=M[(pK^-\pi^+)\pi^+]-M(pK^-\pi^+)+2.286\gevcc\,.
\end{equation}
An extended unbinned maximum likelihood fit is used in the invariant mass range $2.92\gevcc\stackrel{<}{_\sim}M_{\Xi_c}\stackrel{<}{_\sim}3.14\gevcc$.
A scatter plot of the $M_{\Sigma_c}$ vs.~$M_{\Xi_c}$ fit range is shown in Figure~\ref{fg:2D}.
The probability density function (PDF) used to fit the data is divided into four types of components.
One category is used to fit the non-resonant combinatoric background.
Another category is used to fit the combinatoric background with $\Sigma_c(2455)^{++}$ and $\Sigma_c(2520)^{++}$ resonances.
The remaining two categories are used to fit the $\Xi_c(2980)^+$ and $\Xi_c(3077)^+$ signals with and without $\Sigma_c(2455)^{++}$ and $\Sigma_c(2520)^{++}$ resonances.
Each PDF component is described below.

\begin{figure}\begin{center}
\begin{minipage}[t]{0.9\textwidth}
\begin{center}
\includegraphics[width=1.0\textwidth]{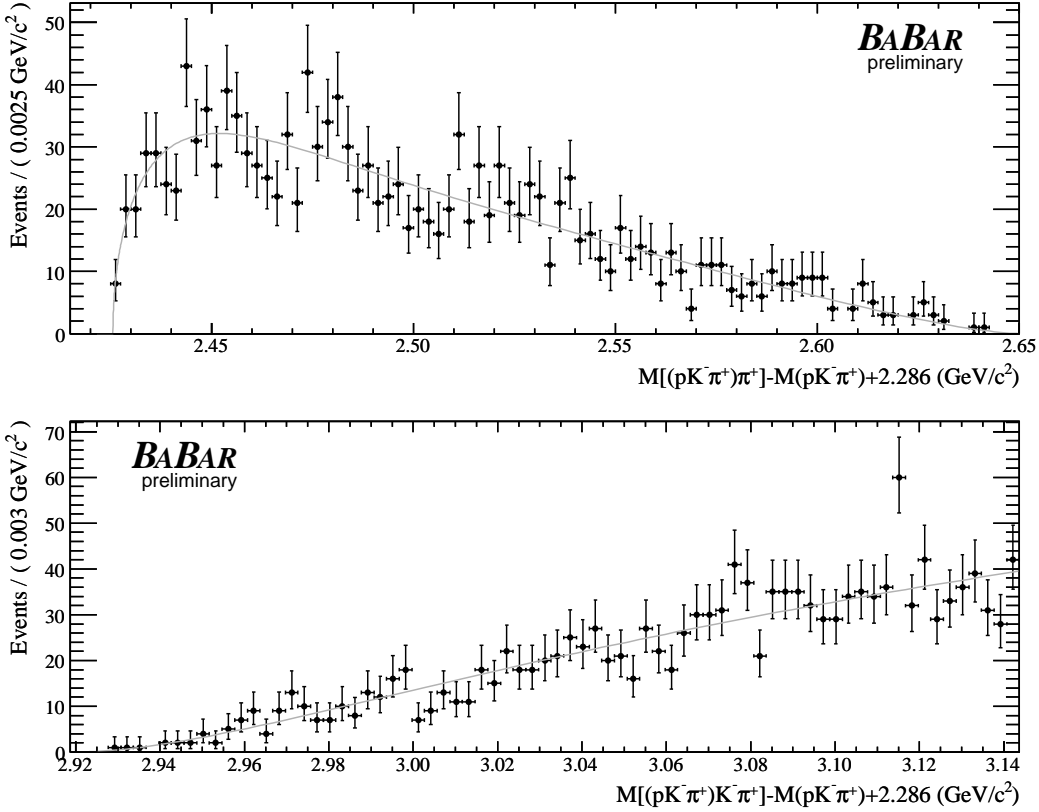}
\caption{Projections onto the mass variables $M_{\Sigma_c}$ (upper) and $M_{\Xi_c}$ (lower) for the $M(pK^-\pi^+)$ sideband sample (points with error bars) and a fitted PDF (curves) described in the text (Equation 3).}
\label{fg:sbbg}
\end{center}
\end{minipage}\end{center}
\end{figure}

A $M(pK^-\pi^+)$ sideband sample around the $\Lambda_c^+$ mass ($20\mevcc<|M(pK^-\pi^+)-2286\mevcc|<40\mevcc$) and a wrong-sign $\Lambda_c^+K^+\pi^-$ data sample are used to establish the non-resonant and resonant background PDF parameterizations, respectively.
Projections of the fit to $M_{\Xi_c}$ and $M_{\Sigma_c}$ in the $M(pK^-\pi^+)$ sideband sample are shown in Figure~\ref{fg:sbbg}.
The PDF used to fit these sideband data is proportional to a threshold function in $M_{\Xi_c}$ ($T(M_{\Xi_c})$) and the sum of two threshold functions in $M_{\Sigma_c}$ ($T(M_{\Sigma_c})$):
\begin{equation}
T(M_{\Xi_c})\times[(1-k)T_a(M_{\Sigma_c})+kT_b(M_{\Sigma_c})]\,,
\end{equation}
where $k$ is a free parameter and the subscripts $a$ and $b$ indicate two instances of the same functional form. These threshold functions are of the form
\begin{equation}
T(x)=x\left[ -1+ \left( \frac{x}{t} \right)^{2} \right]^{1/2}\exp\left[ -p+p\left( \frac{x}{t} \right)^{2}\right]\,,
\end{equation}
where $x$ is the mass variable in which there is a minimum kinematic threshold, $t$ is the mass value of the threshold, and $p$ is a free shape parameter in the fit.
For $T(M_{\Sigma_c})$, the threshold $t$ is a constant 2425.6\mevcc.
For $T(M_{\Xi_c})$, the threshold is dependent on $M_{\Sigma_c}$ through the relation $t=M_{\Sigma_c}+m_K$, where $m_K$ is the $K^+$ mass.

\begin{figure}\begin{center}
\begin{minipage}[t]{0.9\textwidth}
\begin{center}
\includegraphics[width=1.0\textwidth]{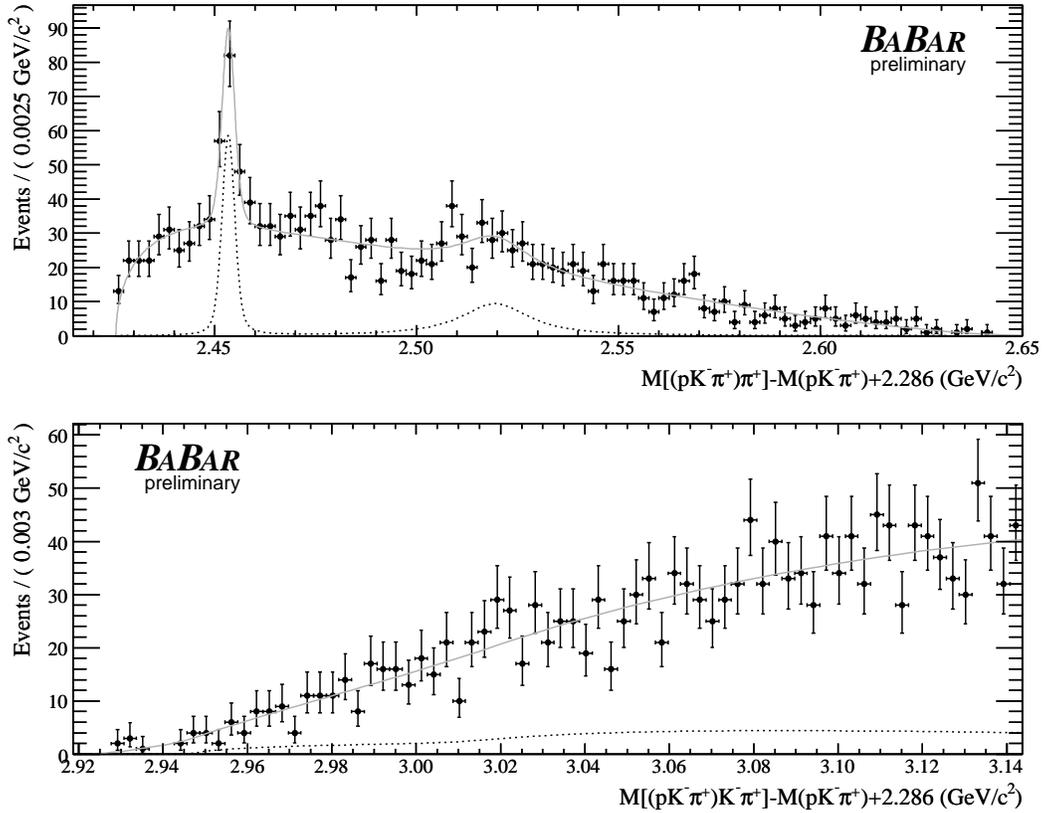}
\caption{Projections onto the mass variables $M_{\Sigma_c}$ (upper) and $M_{\Xi_c}$ (lower) of the wrong-sign $\Lambda_c^+K^+\pi^-$ data sample (points with error bars) and a fitted PDF (curves) described in the text (Equations~3 and 5). The solid curves represent the total fit PDF. The dotted curves represent the fitted resonant combinatoric background PDF component.}
\label{fg:wsbg}
\end{center}
\end{minipage}\end{center}
\end{figure}

Projections of the fit to $M_{\Xi_c}$ vs.~$M_{\Sigma_c}$ in the wrong-sign data sample are shown in Figure~\ref{fg:wsbg}.
The PDF used to fit this wrong-sign sample has two components.
One component is the same as the PDF used to fit the sideband data (Equation 3).
The other component fits $\Sigma_c(2455)^{0}$ and $\Sigma_c(2520)^{0}$ resonances in $M_{\Sigma_c}$ with non-relativistic Breit-Wigner shapes convolved with Gaussian resolution functions (also known as Voigtian line shapes, $V(M_{\Sigma_c})$), times a two-body phase-space function $F_2(M_{\Sigma_c})$.
This second component is proportional to
\begin{equation}
[ (1-r)V_a(M_{\Sigma_c})+rV_b(M_{\Sigma_c}) ]\times F_2(M_{\Sigma_c})\times T(M_{\Xi_c}) \,,
\end{equation}
where $r$ is the ratio of the fitted number of candidates in the $\Sigma_c(2520)^{0}$ resonance to the total fitted number of candidates in both intermediate resonances,
the subscripts $a$ and $b$ indicate two instances of the same Voigtian functional form (one for $\Sigma_c(2455)^0$ and one for $\Sigma_c(2520)^0$),
and $T(M_{\Xi_c})$ has the same shape parameter as $T(M_{\Xi_c})$ in Equation~3.
 The two-body phase-space function is
\begin{equation}
F_2(M_{\Sigma_c})=\frac{[(M_{\Sigma_c}^2-(m_{\Lambda_c}+m_{\pi})^2)(M_{\Sigma_c}^2-(m_{\Lambda_c}-m_{\pi})^2)]^{1/2}}{2M_{\Sigma_c}}\,,
\end{equation}
where $m_{\Lambda_c}$ is the $\Lambda_c^+$ mass used in Equations~1 and 2 (2286\mevcc), and $m_{\pi}$ is the $\pi^+$ mass. 

For both the wrong-sign sample and the $\Lambda_c^+$ sideband sample, the background is described well by the fitted background PDF.
The same PDF functional form used for the wrong-sign sample is used to fit the background candidates in the right-sign data sample where the intermediate resonances are now the $\Sigma_c(2455)^{++}$ and the $\Sigma_c(2520)^{++}$.

The signal PDF components used for both the $\Xi_c(2980)^+$ and $\Xi_c(3077)^+$ non-resonant decays are proportional to
\begin{equation}
F_3(M_{\Xi_c},M_{\Sigma_c})\times V(M_{\Xi_c})\,,
\end{equation}
where $F_3(M_{\Xi_c},M_{\Sigma_c})$ is a three-body phase-space function and $V(M_{\Xi_c})$ is a Voigtian function in $M_{\Xi_c}$.
The three-body phase space function is
\begin{equation}
F_3(M_{\Xi_c},M_{\Sigma_c})=\frac{[(M_{\Xi_c}^2-(M_{\Sigma_c}+m_K)^2)(M_{\Xi_c}^2-(M_{\Sigma_c}-m_K)^2)]^{1/2}}{2M_{\Xi_c}}\times F_2(M_{\Sigma_c})\,,
\end{equation}
where $m_K$ is the $K^+$ mass.

The signal PDF components used for both the $\Xi_c(2980)^+$ and $\Xi_c(3077)^+$ decaying through intermediate resonances are proportional to
the three-body phase-space function $F_3(M_{\Xi_c},M_{\Sigma_c})$ times a Voigtian function in $M_{\Xi_c}$ 
and two Voigtian functions in $M_{\Sigma_c}$:
\begin{equation}
F_3(M_{\Xi_c},M_{\Sigma_c})\times V(M_{\Xi_c})\times [(1-r')V_a(M_{\Sigma_c})+r'V_b(M_{\Sigma_c})] \,,
\end{equation}
where the functions $V_a(M_{\Sigma_c})$ and $V_b(M_{\Sigma_c})$ share the same free parameters as those in the background PDF components,
and $r'$ is an independent ratio parameter used for the signal PDF component.
The $\Xi_c(2980)^+$ cannot decay to $\Sigma_c(2520)^{++}K^-$. In this case, the value of $r'$ is fixed to zero.

Signal-MC samples are used to determine the detector resolution in $M_{\Xi_c}$ and $M_{\Sigma_c}$.
The measured resolutions for $\Xi_c(2980)^+$ and $\Xi_c(3077)^+$ are $(1.6\pm0.1)\mevcc$ and $(2.0\pm0.1)\mevcc$, respectively.
The measured resolutions for $\Sigma_c(2455)^{++}$ and $\Sigma_c(2520)^{++}$ are $(1.3\pm0.2)\mevcc$ and $(1.8\pm0.1)\mevcc$, respectively.

In the fits to data, the simulated detector resolutions are used as fixed parameters in the Voigtian line shapes. Values of kinematic thresholds are also fixed.
All other parameters are free in the fits.  

\section{SYSTEMATIC STUDIES}
\label{sec:Systematics}

Several sources of systematic uncertainty are investigated and quantified.
The results are summarized in Tables~\ref{tb:sys} and \ref{tb:sysB}.

The values of the fixed resolution parameters are changed to determine the effect on the measured mass differences, widths, and yields.
The widths of all convolved resolution-Gaussians are increased and decreased by 10\% in two additional fits to the data.
The largest changes to the mass difference, width, and yield for the $\Xi_c(2980)^+$ signal are $+0.09\mevcc$, $-0.5\mev$ and $-1\%$, respectively.
The largest changes to the mass difference, width, and yield for the $\Xi_c(3077)^+$ signal are $+0.01\mevcc$, $+0.4\mev$ and $-2.2\%$, respectively.
The magnitudes of these changes in fitted values are used as symmetric systematic errors.
Systematic errors for resonant and non-resonant yields are similarly calculated.

In order to evaluate systematic errors due to the shapes of the threshold PDF components, the exponent (1/2) is allowed to be a free parameter in the fit:
\begin{equation}
\left[ -1+\left(\frac{x}{t}\right)^2\right]^{1/2}\rightarrow\left[ -1+\left(\frac{x}{t}\right)^2\right]^{q}\,,
\end{equation}
where $q$ is the new free parameter.
There is one new parameter $q$ used for the threshold components in $M_{\Xi_c}$ and a second for the threshold components in $M_{\Sigma_c}$.
The mass difference, width, and yield for the $\Xi_c(2980)^+$ signal change by $+0.9\mevcc$, $-1.1\mev$, and $-10\%$, respectively.
The mass difference, width, and yield for the $\Xi_c(3077)^+$ signal change by $+0.0004\mevcc$, $+0.17\mev$, and $+0.5\%$, respectively.
The magnitudes of these changes in fitted values are used as symmetric systematic errors.
Systematic errors for resonant and non-resonant yields are similarly calculated.

The $\Xi_c(2980)^+$ and $\Xi_c(3077)^+$ signals are fit with non-relativistic Breit-Wigner shapes convolved with Gaussian resolution functions in the $M_{\Xi_c}$ variable.
We check that there are no substantial errors due to our choice of signal shape.
This is done by fitting the $\Xi_c(2980)^+$ and $\Xi_c(3077)^+$ distributions in signal-MC samples with non-relativistic Breit-Wigner and relativistic S-wave Breit-Wigner shapes convolved with Gaussian resolution functions.
The largest fractional differences found in the resultant mass difference, width, and yield are 0.019\%, 0.018\%, and 0.20\%, respectively. 
These fractional changes are used as symmetric systematic errors and are converted into magnitudes in Tables~\ref{tb:sys} and \ref{tb:sysB}.

The phase-space functions $F_2(M_{\Sigma_c})$ and $F_3(M_{\Xi_c},M_{\Sigma_c})$ are not convolved with the resolution functions in the $M_{\Xi_c}$ or $M_{\Sigma_c}$ variables.
A systematic error due to this PDF inaccuracy is quantified by shifting the $M_{\Xi_c}$ variable in $F(M_{\Xi_c},M_{\Sigma_c})$ by $-2.0\mevcc$, refitting the data, and taking changes in measured quantities as symmetric systematic errors.
Similarly, the $M_{\Sigma_c}$ variable in $F(M_{\Xi_c},M_{\Sigma_c})$ and $F(M_{\Sigma_c})$ is shifted by $-1.3\mevcc$.
Summing these systematic errors in quadrature, the systematic uncertainties for
the $\Xi_c(2980)^+$ mass difference, width, and yield are $\pm0.33\mevcc$, $\pm0.6\mev$ and $\pm13\%$, respectively.
The systematic uncertainties for the $\Xi_c(3077)^+$ mass difference, width, and yield are $\pm0.006\mevcc$, $\pm0.21\mev$ and $\pm5.3\%$, respectively.
Systematic errors for resonant and non-resonant yields are similarly calculated.

Measurements of particle mass with the \babar\ detector have systematic errors associated with SVT alignment, detector angular dependencies, energy-loss corrections, the solenoidal magnetic field, and material magnetization. These systematic errors were extensively studied for \babar's precision measurement of the $\Lambda_c^+$ mass~\cite{lc} and were determined to contribute $\pm0.14\mevcc$ total systematic error to the $\Lambda_c^+$ mass measurement.
The decay mode utilized in the $\Lambda_c^+$ mass measurement ($\Lambda K_S^0K^+$) and the decay mode used in this analysis ($\Lambda_c^+K^-\pi^+$) have similar $Q$-values, where the $Q$-value for a decay $a\rightarrow b+c+\ldots$ is defined as $Q=m_a-m_b-m_c-\ldots$. These similar $Q$-values, along with our more stringent requirement on candidate momentum, lead us to believe that $\pm0.14\mevcc$ is a conservative estimate for the systematic error from detector effects in this analysis.

\begin{table}
\caption[Systematic errors.]{\label{tb:sys} Systematic errors on $\Xi_c(2980)^+$ and $\Xi_c(3077)^+$ mass differences, widths, and yields due to uncertainties in signal resolution, phase-space kinematic suppression, and the background PDF parameterization. The systematic errors from each source are added in quadrature.}
\begin{center}
\begin{tabular}{cccc} \hline
                                 & Mass Difference (\mevcc) & Width (\mev)  & Yield (\%) \\ \hline\hline
$\Xi_c(2980)^+$ \\
Resolution                       & $\pm0.09$                & $\pm0.5$      & $\pm1$    \\
Background Parameters            & $\pm0.88$                & $\pm1.1$      & $\pm10$   \\
Breit-Wigner Shape               & $\pm0.13$                & $<0.05$       & $<0.5$    \\
Phase-Space                      & $\pm0.33$                & $\pm0.6$      & $\pm13$   \\
Detector Effects                 & $\pm0.14$                & |             & |         \\ \hline
Total                            & $\pm0.96$                & $\pm1.3$      & $\pm16$   \\ \hline
$\Xi_c(3077)^+$ \\
Resolution                       & $\pm0.01$                & $\pm0.44$     & $\pm2.2$   \\
Background Parameters            & $<0.005$                 & $\pm0.17$     & $\pm0.5$   \\ 
Breit-Wigner Shape               & $\pm0.15$                & $<0.005$      & $\pm0.2$   \\
Phase-Space                      & $\pm0.06$                & $\pm0.21$     & $\pm5.3$   \\
Detector Effects                 & $\pm0.14$                & |             & |          \\ \hline
Total                            & $\pm0.21$                & $\pm0.52$     & $\pm5.8$   \\ \hline \hline
\end{tabular}
\end{center}
\end{table}

\begin{table}
\caption[Additional systematic errors.]{\label{tb:sysB} Systematic errors on $\Xi_c(2980)^+$ and $\Xi_c(3077)^+$ resonant and non-resonant decay yields due to uncertainties in signal resolution, phase-space kinematic suppression, and the background PDF parameterization. The systematic errors from each source are added in quadrature.}
\begin{center}
\begin{tabular}{cccccc} \hline
                                                   & Resolution   & Phase-Space   & Breit-Wigner   & Background   & Total       \\ \hline\hline
$\Xi_c(2980)^+\rightarrow\Sigma_c(2455)^{++}K^-$   & $\pm2.3\%$   & $\pm1.4\%$    & $\pm0.2\%$     & $\pm2.2\%$   & $\pm3.5\%$  \\
$\Xi_c(2980)^+\rightarrow\Lambda_c^+K^-\pi^+$      & $\pm1.9\%$   & $\pm24.9\%$   & $\pm0.2\%$     & $\pm15.7\%$  & $\pm29.5\%$ \\
$\Xi_c(3077)^+\rightarrow\Sigma_c(2455)^{++}K^-$   & $\pm3.1\%$   & $\pm1.6\%$    & $\pm0.2\%$     & $\pm2.2\%$   & $\pm4.1\%$  \\
$\Xi_c(3077)^+\rightarrow\Sigma_c(2520)^{++}K^-$   & $\pm2.3\%$   & $\pm3.4\%$    & $\pm0.2\%$     & $\pm5.7\%$   & $\pm7.0\%$  \\
$\Xi_c(3077)^+\rightarrow\Lambda_c^+K^-\pi^+$      & $\pm10.6\%$  & $\pm42.0\%$   & $\pm0.2\%$     & $\pm15.7\%$  & $\pm46.1\%$ \\ \hline \hline
\end{tabular}
\end{center}
\end{table}

\section{PHYSICS RESULTS}
\label{sec:Physics}

\begin{figure}\begin{center}
\begin{minipage}[t]{0.9\textwidth}
\begin{center}
\includegraphics[width=1.0\textwidth]{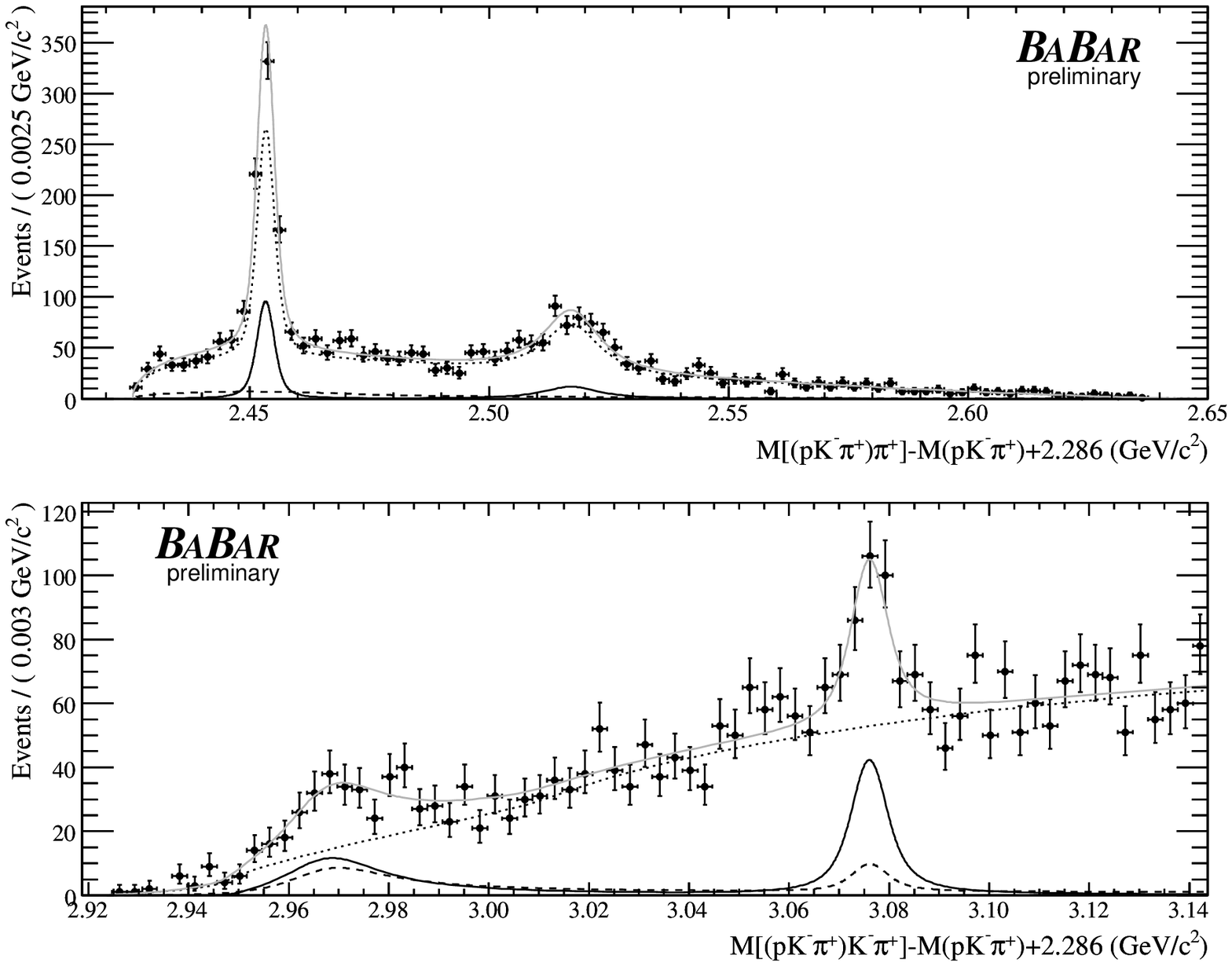}
\caption{Projections onto the mass variables $M_{\Sigma_c}$ (upper) and $M_{\Xi_c}$ (lower) for the data (points with error bars) and the fitted two-dimensional PDF (curves). The solid gray curves represent the total fit PDF. The dotted curves represent the sum of the background components with no intermediate resonances and with the $\Sigma_c(2455)^{++}$ and $\Sigma_c(2520)^{++}$ intermediate resonances. The solid dark curves represent the sum of the $\Xi_c(2980)^+\rightarrow\Sigma_c(2455)^{++}K^-$ signal component, the $\Xi_c(3077)^+\rightarrow\Sigma_c(2455)^{++}K^-$ signal component, and the $\Xi_c(3077)^+\rightarrow\Sigma_c(2520)^{++}K^-$ signal component. The dashed curves represent the sum of the $\Xi_c(2980)^+\rightarrow\Lambda_c^+K^-\pi^+$ and $\Xi_c(3077)^+\rightarrow\Lambda_c^+K^-\pi^+$ signal components.}
\label{fg:fit}
\end{center}
\end{minipage}\end{center}
\end{figure}

\begin{figure}\begin{center}
\begin{minipage}[t]{0.9\textwidth}
\begin{center}
\includegraphics[width=1.0\textwidth]{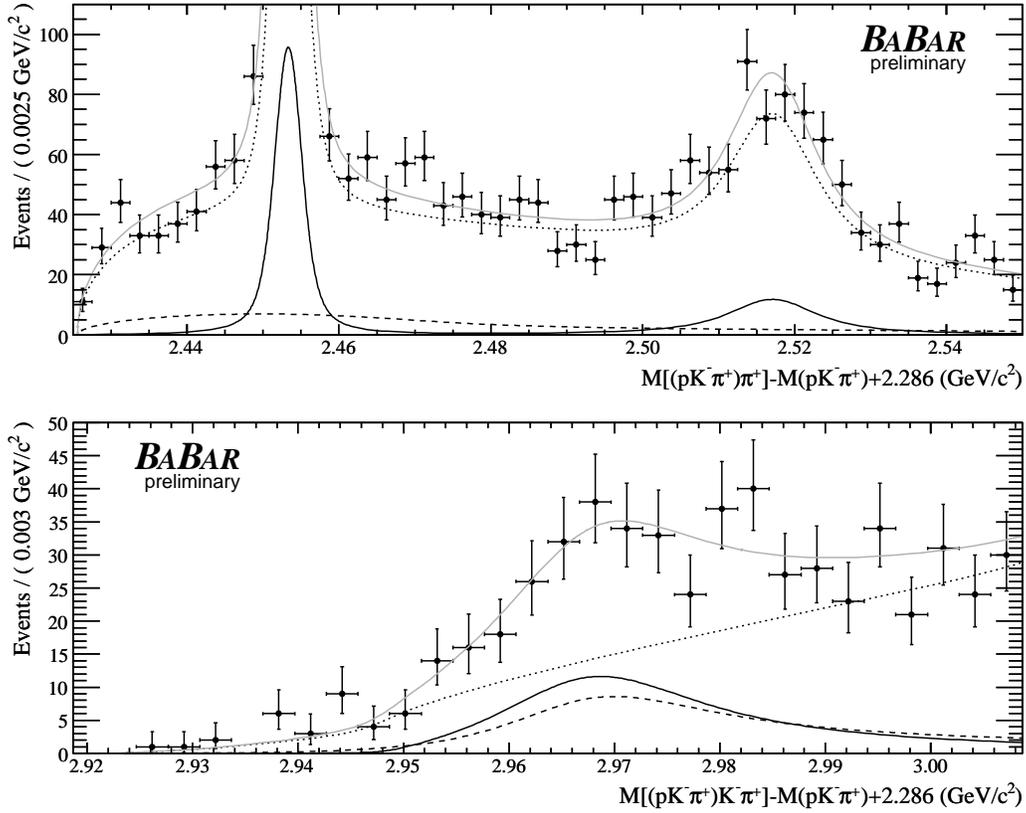}
\caption{Projections onto the mass variables $M_{\Sigma_c}$ (upper) and $M_{\Xi_c}$ (lower) for the data (points with error bars) and the fitted PDF (curves), magnified to further illustrate the different PDF components. The curves are the same as in Figure~\ref{fg:fit}. The solid gray curves represent the total fit PDF. The dotted curves represent the sum of the background components with no intermediate resonances and with the $\Sigma_c(2455)^{++}$ and $\Sigma_c(2520)^{++}$ intermediate resonances. The solid dark curves represent the sum of the $\Xi_c(2980)^+\rightarrow\Sigma_c(2455)^{++}K^-$ signal component, the $\Xi_c(3077)^+\rightarrow\Sigma_c(2455)^{++}K^-$ signal component, and the $\Xi_c(3077)^+\rightarrow\Sigma_c(2520)^{++}K^-$ signal component. The dashed curves represent the sum of the $\Xi_c(2980)^+\rightarrow\Lambda_c^+K^-\pi^+$ and $\Xi_c(3077)^+\rightarrow\Lambda_c^+K^-\pi^+$ signal components.}
\label{fg:fit2}
\end{center}
\end{minipage}\end{center}
\end{figure}

The data is fit to determine $\Xi_c(2980)^+$ and $\Xi_c(3077)^+$ signal widths, mass differences with respect to $\Lambda_c^+$,
and total yields, as well as yields for resonant and non-resonant decays.
Figure~\ref{fg:fit} shows projections of the data and the fit results.
Figure~\ref{fg:fit2} shows the same projections of the data and the fit, but with regions magnified to further illustrate the individual PDF components.

In order to determine the statistical significance of the $\Xi_c(2980)^+$ and $\Xi_c(3077)^+$ signals,
fits to the data are performed without each of the signal components for the $\Xi_c(2980)^+$ and the $\Xi_c(3077)^+$.
The maximum log likelihood for the fit decreases by 28.9 units when the $\Xi_c(2980)^+$ signal PDF is excluded from the fit.
This decrease in maximum log likelihood, with the joint estimation of three parameters (mass, width, and yield), corresponds to a $7.0\,\sigma$ significance for the $\Xi_c(2980)^+$ signal.
The maximum log likelihood decreases by 42.5 units when the $\Xi_c(3077)^+$ signal is excluded from the fit.
This decrease in the maximum log likelihood, again with the joint estimation of three parameters, corresponds to a significance of $8.6\,\sigma$.

The measured mass differences with respect to the $\Lambda_c^+$ are
\begin{displaymath}
M(\Xi_c(2980)^+)-M(\Lambda_c^+)= (680.6\pm1.9\pm1.0)\mevcc\,{\rm, and}
\end{displaymath}\begin{displaymath}
M(\Xi_c(3077)^+)-M(\Lambda_c^+)= (790.0\pm0.7\pm0.2)\mevcc\,.
\end{displaymath}
The masses, widths, yields, and significances for the $\Xi_c(2980)^+$ and $\Xi_c(3077)^+$ are listed in Table~\ref{tb:results}.
The $\Xi_c(2980)^+$ and $\Xi_c(3077)^+$ masses are calculated from their mass differences with respect to the $\Lambda_c^+$ by adding the $\Lambda_c^+$ mass $(2286.46\pm0.14)\mevcc$ as measured by \babar~\cite{lc}.
The total yields for the $\Xi_c(2980)^+$ and $\Xi_c(3077)^+$ signals are the combined yields from their resonant and non-resonant decays,
and take into account the correlations due to the shared mean mass-difference and width parameters.
Masses, widths, yields, and significances, as measured by the Belle Collaboration in the $\Lambda_c^+K^-\pi^+$ final state~\cite{Belle1,Belle2}, are also listed for comparison. The quoted Belle significance are calculated assuming the estimation of one parameter.

The yields and significances for the separate resonant and non-resonant decays are listed in Table~\ref{tb:yields}.
The significances for the resonant and non-resonant decays of the $\Xi_c(2980)^+$ and $\Xi_c(3077)^+$ are each calculated separately
with the same method as for the full signal but with the yield being the only parameter estimated.
We find that the signal for the resonant decay $\Xi_c(2980)^+\rightarrow\Sigma_c(2455)^{++}K^-$ has a $4.9\,\sigma$ significance.
The signal for the non-resonant decay $\Xi_c(2980)^+\rightarrow\Lambda_c^+K^-\pi^+$ has a $4.1\,\sigma$ significance.
We find that the signal for the resonant decay $\Xi_c(3077)^+\rightarrow\Sigma_c(2455)^{++}K^-$ has a $5.8\,\sigma$ significance,
and the signal for $\Xi_c(3077)^+\rightarrow\Sigma_c(2520)^{++}K^-$ has a $4.6\,\sigma$ significance.
The signal for the non-resonant decay $\Xi_c(3077)^+\rightarrow\Lambda_c^+K^-\pi^+$ has a $1.4\,\sigma$ significance.

\begin{table}
\caption{\label{tb:results} Comparison of masses, widths, yields, and significances for $\Xi_c(2980)^+$ and $\Xi_c(3077)^+$, measured by \babar\ and Belle in the $\Lambda_c^+K^-\pi^+$ final state. The quoted Belle significance are calculated assuming the estimation of one parameter; the \babar\ significances are calculated for the joint estimation of three parameters (mass, width and yield).}
\begin{center}
\begin{tabular}{rcccc} \hline
                        & Mass (\mevcc)           & Width (\mev)          & Yield (Events)     & Significance     \\ \hline\hline
\babar\ $\Xi_c(2980)^+$ & $2967.1\pm1.9\pm1.0$    & $23.6\pm2.8\pm1.3$    & $284\pm45\pm46$    & $7.0\,\sigma$    \\
Belle $\Xi_c(2980)^+$   & $2978.5\pm2.1\pm2.0$    & $43.5\pm7.5\pm7.0$    & $405\pm51$         & $6.3\,\sigma$    \\
\babar\ $\Xi_c(3077)^+$ & $3076.4\pm0.7\pm0.3$    & $6.2\pm1.6\pm0.5$     & $204\pm35\pm12$    & $8.6\,\sigma$    \\
Belle $\Xi_c(3077)^+$   & $3076.7\pm0.9\pm0.5$    & $6.2\pm1.2\pm0.8$     & $326\pm40$         & $9.7\,\sigma$    \\ \hline\hline
\end{tabular}
\end{center}
\end{table}

\begin{table}
\caption{\label{tb:yields} Yields and significances for the separate resonant and non-resonant decays.}
\begin{center}
\begin{tabular}{lcc} \hline
                                                     & Yield (Events)     & Significance     \\ \hline\hline
$\Xi_c(2980)^+\rightarrow\Sigma_c(2455)^{++}K^-$     & $132\pm31\pm5$     & $4.9\,\sigma$    \\
$\Xi_c(2980)^+\rightarrow\Lambda_c^+K^-\pi^+$        & $152\pm37\pm45$    & $4.1\,\sigma$    \\ \hline
$\Xi_c(3077)^+\rightarrow\Sigma_c(2455)^{++}K^-$     & $87\pm20\pm4$      & $5.8\,\sigma$    \\
$\Xi_c(3077)^+\rightarrow\Sigma_c(2520)^{++}K^-$     & $82\pm23\pm6$      & $4.6\,\sigma$    \\
$\Xi_c(3077)^+\rightarrow\Lambda_c^+K^-\pi^+$        & $35\pm24\pm16$     & $1.4\,\sigma$    \\ \hline\hline
\end{tabular}
\end{center}
\end{table}
                                                                                                                                                            
\section{SUMMARY}
\label{sec:Summary}
We analyze $315.7\invfb$ of data collected with the \babar\ detector and search for $\Xi_c(2980)^+\rightarrow\Lambda_c^+K^-\pi^+$ and $\Xi_c(3077)^+\rightarrow\Lambda_c^+K^-\pi^+$.
A significant signal is found for the non-resonant decay of $\Xi_{c}(2980)^+$ to $\Lambda_c^+K^-\pi^+$ as well as for a resonant decay through $\Sigma_c(2455)^{++}K^-$.
Significant signals are found for resonant decays of $\Xi_c(3077)^+$ to $\Lambda_c^+K^-\pi^+$ through the intermediate states $\Sigma_c(2455)^{++}K^-$ and $\Sigma_c(2520)^{++}K^-$.
We find only a small indication that $\Xi_c(3077)^+$ decays non-resonantly to $\Lambda_c^+K^-\pi^+$. 
Our measured values of $\Xi_c(3077)^+$ mass and width are consistent with results from Belle~\cite{Belle1,Belle2}.
However, our measured values of $\Xi_{c}(2980)^+$ mass and width are significantly lower and narrower, respectively, than those measured by Belle.
This may be due to our use of a two-dimensional fit and phase-space considerations.

\section{ACKNOWLEDGMENTS}
\label{sec:Acknowledgments}

We are grateful for the 
extraordinary contributions of our \pep2\ colleagues in
achieving the excellent luminosity and machine conditions
that have made this work possible.
The success of this project also relies critically on the 
expertise and dedication of the computing organizations that 
support \babar.
The collaborating institutions wish to thank 
SLAC for its support and the kind hospitality extended to them. 
This work is supported by the
US Department of Energy
and National Science Foundation, the
Natural Sciences and Engineering Research Council (Canada),
Institute of High Energy Physics (China), the
Commissariat \`a l'Energie Atomique and
Institut National de Physique Nucl\'eaire et de Physique des Particules
(France), the
Bundesministerium f\"ur Bildung und Forschung and
Deutsche Forschungsgemeinschaft
(Germany), the
Istituto Nazionale di Fisica Nucleare (Italy),
the Foundation for Fundamental Research on Matter (The Netherlands),
the Research Council of Norway, the
Ministry of Science and Technology of the Russian Federation, 
Ministerio de Educaci\'on y Ciencia (Spain), and the
Particle Physics and Astronomy Research Council (United Kingdom). 
Individuals have received support from 
the Marie-Curie IEF program (European Union) and
the A. P. Sloan Foundation.

\end{document}